**Compound-specific isotope analysis (CSIA). Application to archaeology, biomedical sciences, biosynthesis, environment, extraterrestrial chemistry, food science, forensic science, humic substances, microbiology, organic geochemistry, soil science and sport.**

Eric LICHTFOUSE

Soil and Environment Laboratories, INRA-ENSAIA/INPL, BP 172, 54505 Vandoeuvre-les-Nancy, France.

<Eric.Lichtfouse@ensaia.inpl-nancy.fr>

**ABSTRACT**

The isotopic composition, e.g. $^{14}C/^{12}C$, $^{13}C/^{12}C$, $^{2}H/^{1}H$, $^{15}N/^{14}N$ and $^{18}O/^{16}O$, of matter elements is heterogeneous. It is ruled by physical, chemical and biological mechanisms. Isotopes thus be enable to follow the fate of mineral and organic compounds during biogeochemical transformations. The determination of the isotopic composition of organic substances occurring at trace level into very complex mixtures such as sediments, soils and blood, has been made possible during the last 20 years due to the rapid development of molecular level isotopic techniques. After a brief glance at pioneer studies revealing isotopic breakthroughs the molecular and intramolecular levels, this paper reviews selected applications of compound-specific isotope analysis in various scientific fields.

**INTRODUCTION**

Most stable isotopic research relies on two concepts. First, natural and artificial chemical reactions fractionate isotopes, thus leading to the occurrence of various organic and inorganic materials having different isotopic compositions. Isotopes can thus record biogeochemical changes. Second, isotopes can be used as natural or artificial tracers to follow the behaviour of organic molecules in complex media such as living organisms and ecosystems. The determination of $^{13}C/^{12}C$ ratios[1] of natural organic matter has been restricted for a long time to the analysis of bulk samples[2]. Then, the development of gas chromatography coupled to a combustion furnace then to an isotope ratio mass spectrometer[3-5] (GC-C-IRMS, **Figure**) has allowed the analysis of individual substances occurring at trace levels in very complex mixtures. It thus opened new research fields in various scientific areas[6-8] such as organic geochemistry[9,10], food science[11], medecine[12], nutrition[13], pharmacy[14], sport[15], phytochemistry[16,17], archaeology[18], soil science[19,20], environment[21-25], humic substances[26-28], extraterrestrial science[29] and forensic science[30,31]. $^{15}N/^{14}N$ and $^{2}H/^{1}H$ ratios of individual molecules have also been measured using a GC coupled to an IRMS[32-35]. Besides, since GC is limited to the study of volatile substances, another hyphenated technique has been developed, coupling liquid chromatography to an isotopic ratio mass spectrometer via a combustion furnace (LC-C-IRMS), thus enabling high-molecular, polar, or thermally sensible molecules to be analysed[36,37]. Further, determination of $^{2}H/^{1}H$ and $^{13}C/^{12}C$ ratios of each atomic site of a pure substance can be performed by site-specific natural isotope fractionation nuclear magnetic resonance (SNIF-NMR)[38-40]. $^{14}C$ dating of individual substances occurring in complex mixtures has been recently achieved using preparative-GC to isolate pure lipids[41]. Hereafter, in the Early Days section, several examples will show how some pioneers have been able to determine $^{13}C/^{12}C$ ratios of individual substances, and even of individual atomic sites, before GC-C-IRMS has been available. Then, selected applications of GC-C-IRMS in various scientific fields will be given.

FIGURE

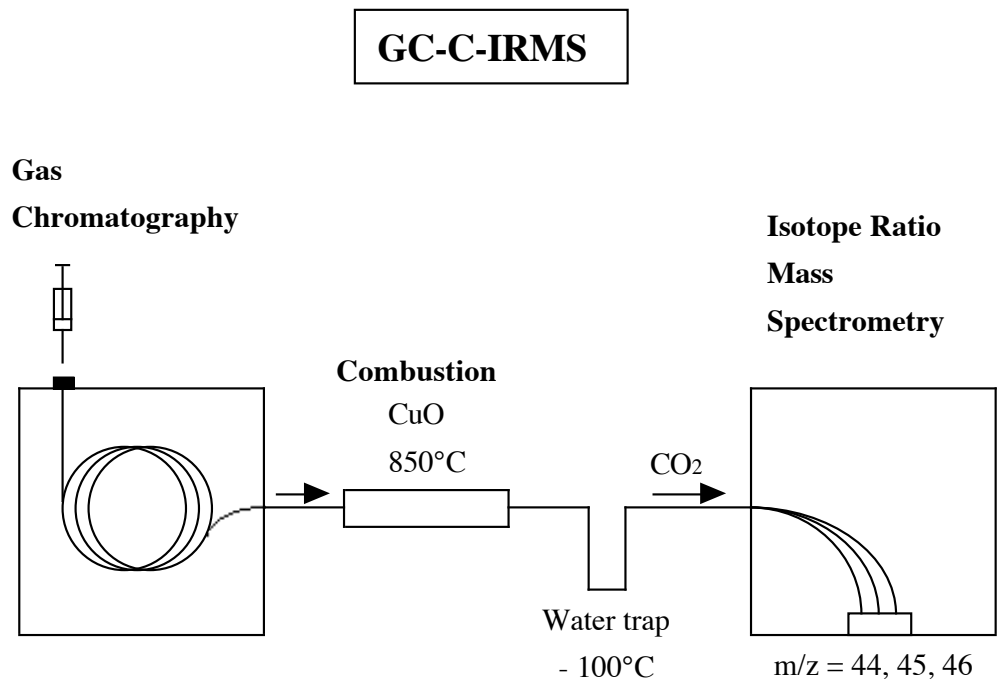

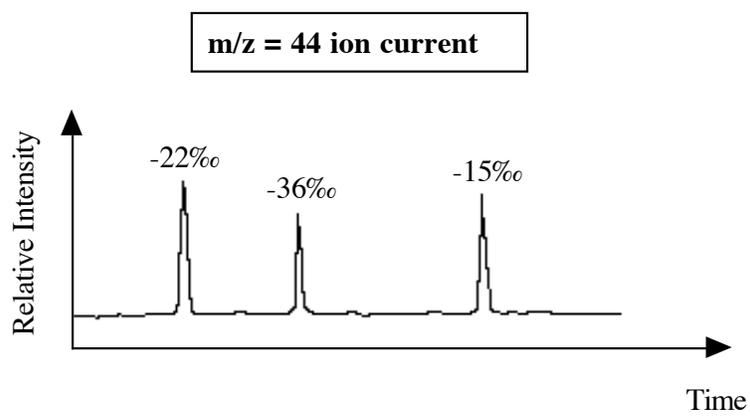

**EARLY DAYS**

Before the advent of hyphenated techniques allowing the on-line purification of complex mixtures, off-line isotopic analysis relied on the availability of a pure substance in reasonable amounts, from about 10µg to 10mg, for quartz tube CuO combustion followed by $CO_2$ transfer into the mass spectrometer. Therefore the analysis of individual compounds occurring in complex media such as living organisms and sediments required careful analytical fractionation to yield pure substances. Here, a milestone has been set by Abelson and Hoering[42] who isolated individual amino acids from algal and bacterial cultures by several analytical steps including ion-exchange chromatography. They found that amino acids and lipids were respectively $^{13}C$-enriched and $^{13}C$-depleted relative to the total organic carbon. Moreover, after ninhydrin decarboxylation of amino acids, they showed that carboxyl groups are strongly enriched in carbon 13 relative to the whole molecule, thus making $^{13}C$ investigations a promising tool to study biosynthesis. This non-statistical isotopic composition at each atomic site has also been shown by NMR[43] for algal amino acids grown on $^{13}C$-$CO_2$.

Parker measured $^{13}C/^{12}C$ ratios of individual fatty acids isolated from algae and grasses by preparative GC[44,45]. The results show that lipids, e.g. fatty acids, were $^{13}C$-depleted up to 15‰ relative to the bulk organic carbon. Using the Schmidt decarboxylation[46], Vogler and Hayes showed that the carboxyl group of fatty acids from corn, soybean and algae is enriched in carbon 13[47]. These authors have also set up a method to synthesise fatty acids with known isotope value of the carboxyl carbon[48]. Monson and Hayes cleaved the double bonds of unsaturated fatty acids with ozone in order to determine intramolecular isotopic compositions[49-50]. Meinschein et al. found that the methyl group of acetic acid from cider vinegar is $^{13}C$-depleted relative to the whole molecule, while commercial acetic acid from inorganic synthesis showed the reverse[51]. In close agreement, the carboxyl carbon of acetic acid excreted by the bacterium *Acetobacter suboxydans* grown on ethanol is $^{13}C$-enriched by 3.8‰ relative to the methyl carbon[52]. Rinaldi et al. determined the isotopic composition of individual atomic sites of 3-hydroxy-2-butanone from apple cider vinegar and found that $^{13}C/^{12}C$ ratios increase with increasing oxidation level : ketone, alcohol, then methyl[53]. Remarkably, they write in 1974 that « because the distribution of isotopes within biological compounds are controlled by kinetic and equilibrium effects, intramolecular isotopic analyses may provide a means of defining the equilibria between metabolic pathways and of recognising the molecular remnants of pre-existent organisms in geological samples ».

At a time when the biological origin of petroleum was still a matter of debate, Welte, using preparative GC, measured heterogeneous $^{13}C/^{12}C$ values of individual *n*-alkanes from an Eocene crude petroleum, a possible clue to their biological origin[54,55]. Galimov and Shirinsky reported $^{13}C/^{12}C$ values of various natural organic substances isolated from marine algae, krill zooplancton and lupin plant by thin layer chromatography[56]. These authors proposed a model to calculate intramolecular isotope values. Galimov et al. reported the intramolecular isotope heterogeneity of benzaldehydes, e.g. vanillin, isolated from an Asiatic reed by distillation[57].

**ARCHAEOLOGY**

Isotope analysis of individual substances could be a powerful way to bring some light on the origin and on the history of ancient debris, especially when morphological details are absent. For instance, Evershed et al. showed that $^{13}C/^{12}C$ ratios of lipids preserved in archaeological potsherds are consistent with those of modern cabbage, thus revealing ancient dietary practices[18]. Dudd and Evershed further assessed the use of milk in archaeological economies, using $^{13}C/^{12}C$ ratios of fatty acids extracted from pottery vessels of late Saxon and Iron age[58]. Analysing bone cholesterol from fossil whales, Stott et al. recorded $^{13}C/^{12}C$ ratios within the range expected for the fat of marine mamals[59]. They also showed a similar depletion between cholesterol and collagen for both modern and ancient whale bones, indicating the reliable preservation of the isotope signal. Further, the recent set up of $^{14}C$ dating at the molecular level by Eglinton et al. now enables to date and source individual organic substances occurring in archaeological debris[60,61].

**BIOMEDICAL**

The analysis of stable isotopes at the molecular level for biomedical issues is particularly promising because substances labelled with stable isotopes could ultimately replace the use of radioactive substances. Recent applications of stable isotopes in biomedical fields such as human metabolism[12], nutrition[12,13], and pharmacy[14] have been reviewed. Tissot et al. monitored the metabolic fate of $^{13}C$-glucose in rats and humans by CG-C-IRMS[62]. In a similar way, Khalfallah et al. designed a method to measure plasma lactate after derivatization[63]. Koziet et al. followed the incorporation of $^{13}C$-enriched ethanol into human plasmatic triglycerides[64]. Preston and Slater discussed the respective advantages of IRMS, continuous flow (CF)-IRMS, GC-MS and GC-C-IRMS for the use of amino acid as tracers[65]. Yarasheski et al. monitored by GC-C-IRMS the incorporation of $^{13}C$-leucine into muscle protein of rats[66]. To test the effect of the anti-inflammatory agent isoprofen on acute-phase protein response in patients with colonic cancer, Preston et al. measured the $^{15}N$ enrichment of plasma and tissue glycine after administration of $^{15}N$-glycine[67]. Following oral loads of [$^{13}C$-carboxyl]-isoleucine, Meier-Augenstein et al. followed the $^{13}C$-enrichment of 2-keto-3-methylpentanoic acid in human plasma[68].

Abnormalities in cholesterol metabolism are implicated in the development of cardiovascular disease in humans. Here, Guo et al. studied by GC-C-IRMS the turnover of human plasma cholesterol following oral administration of [3,4-$^{13}C$]-cholesterol[69]. Deruaz et al. reported a comparison of GC-MS, GC-C-IRMS and GC-atomic emission detector (AED) for the analysis of [3,4-$^{13}C$]-progesterone[70]. The turnover of fatty acid can also be studied by GC-C-IRMS. For instance, Goodman and Brenna followed the $^{13}$C-enrichment with time of stearic acid (C$_{18}$) in human plasma following administration of $^{13}$C-stearic acid[71]. Brossard et al. studied the incorporation of polyunsaturated fatty acids into rat lipoproteins and red cells after ingestion of $^{13}$C-labelled triglycerides[72]. Similarly, the incorporation of fatty acids in rat brain has been described by Brookes et al.[73]. Since fatty acids are usually analysed as their methyl ester derivatives, methods to take into account the isotopic value of the added methyl group have been described[74,75]. In order to measure triacylglycerol synthesis rate in humans, Scrimgeour et al. used GC-Py-IRMS to measure the incorporation of D$_2$O into plasma fatty acids[76].

**BIOSYNTHESIS**

Since $^{13}$C/$^{12}$C ratios of biological substances depend upon isotope effects associated with metabolism and biosynthesis, as outlined by Hayes[77], geochemists have studied the extend of isotope fractionation in order to bring some light on the biological sources of sedimentary lipids. Fang et al. reported $^{13}$C/$^{12}$C ratios of fatty acids from hydrocarbon seep mytilids and vestimentiferan[78]. The results suggest isotope fractionation during fatty acid desaturation and elongation. Investigations of plants using C$_3$, C$_4$ and CAM metabolisms[79] showed differences of $^{13}$C/$^{12}$C ratios between *n*-alkanes homologues from the same plant[16,17,80]. Lockheart et al. reported seasonal variations of $^{13}$C/$^{12}$C ratios of *n*-alkanes from tree leaves[81]. Lichtfouse inferred a similar biosynthetic source for long-chain linear alkanes, alcohols and acids in soils based on isotopic signals[82]. Summons et al. found that *n*-fatty acids from the methanotrophic bacterium *Methylococcus capsulatus* were unexpectedly $^{13}$C-enriched compared to polyisoprenoids[83]. Cultures of antartic methanogens revealed isotope effects up to 80‰ between the substrate and phytanyl ethers[84]. Van der Meer reported $^{13}$C-enriched values for lipids from the green sulphur bacterium *Chlorobium limicola*, an organism using the reversed tricarboxylic acid cycle[85]. $^{2}$H/$^{1}$H isotope variations as large as 150‰ have been observed among isoprenoid lipids from a single organism[35]. By incubating a sediment with $^{13}$C-labelled microalgae, Sun followed fatty acid metabolites by gas chromatography coupled to mass spectrometry (GC-MS)[86].

**ENVIRONMENT**

Sherwood Lollar and Abrajano have recently edited a special issue of the journal Organic Geochemistry devoted to the use of compound-specific analysis for environmental issues[21]. Pollution of sediments[22,87,88] and soils[24] by petroleum *n*-alkanes has been inferred using $^{13}$C/$^{12}$C ratios. In soils, the ancient origin of petroleum alkanes was confirmed by $^{14}$C-dating[24]. Lichtfouse and Eglinton reconstructed the fingerprint of the petroleum contamination using a two end-member mixing model[24]. Lichtfouse et al. further showed that short-chain *n*-alkanes from crop soils represent a fossil fuel contamination because these compounds did not show any $^{13}$C/$^{12}$C variation during a 23-year experiment of labelling of soil carbon by maize cultivation[89].

The occurrence of polycyclic aromatic hydrocarbons (PAHs) in the modern environment[90,91] has focused interest because these compounds could exhibit toxicological activity. O'Malley et al. reported $^{13}C/^{12}C$ distinct values for PAHs from various environmental sources such as carsoot, fireplace, sewage and harbour sediments[23]. The production of PAHs by biomass burning did not show significant isotopic fractionation, thus enabling the use of isotope analysis to source sedimentary PAHs derived from fires[92]. Ballentine et al. reported $^{13}C/^{12}C$ values of PAHs and fatty acids from aerosols collected during a field burn of sugar cane[93]. McRae et al. inferred the fossil fuel sources of PAHs from soils and vegetation sampled close to a carbonisation plant[94]. Lichtfouse et al. assessed the fossil fuel origin of PAHs in crop soils using $^{13}C$-labelling with maize, $^{14}C$-dating and biomarker evidence[95]. Richnow et al. monitored the incorporation of [9-$^{13}C$]-anthracene and its metabolites into soil humus[96]. In an isotopic survey of commercial benzene, toluene, ethylbenzene and xylenes, Harrigton et al. observed a wide range of $^{13}C/^{12}C$ values, suggesting applications to qualify and to quantify subsurface processes affecting contaminant concentrations[97]. $^{13}C/^{12}C$ isotope fractionation was observed by Meckenstock et al. during bacterial degradation of toluene[98].

Zeng et al. applied GC-C-IRMS to measure $^{13}C/^{12}C$ ratios of greenhouse gases from biomass burning and exhausts[99]. Biodegradation of petroleum hydrocarbons has been monitored using $^{13}C$, $^{14}C$ and $^{2}D$ contents of $CH_4$ and $CO_2$[100,101]. Rowe and Muehlenbachs proposed an isotopic means to assess the contamination of shallow aquifers by petroleum gases ($C_1$-$C_4$) migrating from deeper leaking wells[102]. Dayan et al. studied the isotopic fractionation of chlorinated ethenes during reductive dehalogenation by metallic iron, a recently developed remediation technology[103]. As shown by Heraty et al., $^{13}C/^{12}C$ investigations of chorinated hydrocarbons, e.g. dichloromethane, could lead to applications in the characterisation and remediation of contaminated sites[104]. For instance, Sherwood Lollar et al. recorded a $^{13}C$ increase from -30‰ to -16‰ during the anaerobic biodegradation of trichloroethylene[105]. A similar trend was observed by Stehmeier et al. during benzene degradation[106]. Bergamaschi et al. found that trihalomethanes formed by reaction of chlorine with plant leachates are $^{13}C$-depleted of 12‰ relative to the whole plant material[107]. Further environmental applications will benefit from the set up by Dias and Freeman of GC-C-IRMS analysis of aqueous volatile organic compounds using solid-phase microextraction (SPME)[108].

**EXTRATERRESTRIAL CHEMISTRY**

A major issue in the analysis of extraterrestrial material such as meteorites is the contamination of samples by terrestrial matter. Here, isotopic analysis is particularly relevant because extraterrestrial compounds have usually sharply different isotopic signatures compared to earth substances. For instance, Epstein et al. found unusually high $^{2}H/^{1}H$ and $^{15}N/^{14}N$ ratios in amino acid and monocarboxylic acid fractions from the Murchison meteorite, thus confirming the extraterrestrial origin of both classes of compounds[109]. Engel et al.[29] further measured by GC-C-IRMS the $^{13}C/^{12}C$ ratios of individual amino acids[110] from the same meteorite. Alanine being not racemic in meteorites, the $^{13}C$-enrichment of its D- and L-enantiomers[110,111] implies that the excess of the L-enantiomer is indigenous rather than from terrestrial contamination. These results suggest that optically active materials were present in the early Solar System before life began. Such findings bring some light on the question of why living systems on Earth came to use almost exclusively left-handed amino acids (L-enantiomers).

**FOOD SCIENCE**

Reviews on carbon isotopes in food technology[112], NMR isotopic analysis[113], GC studies of flavours and fragrances[114] and authentication[115] have been published. Isotopes are mainly used in the food science area to distinguish natural versus synthetic origin of ingredients. Indeed, cheap, synthetic, petroleum-derived compounds have usually different isotope values than their costly, natural counterparts. Several studies highlight the use of both chirospecific analysis and $^{13}C/^{12}C$ determination by CG-C-IRMS for the authentication of fruit lactones[11,116-119]. Schmidt reported $^{13}C/^{12}C$ values of various synthetic and natural compounds such as carvone, almond benzaldehyde, and cinnamon aldehyde[120]. Breaking the molecules into pieces prior isotope analysis, these authors also showed that intramolecular isotope values are heterogeneous. For instance, the aglycone moieties of glycosides is $^{13}C$-depleted relative to the carbohydrate counterpart, as a consequence of different biosynthetic pathways. Braunsdorf et al. further studied $^{13}C/^{12}C$ values of flavour compounds from fruit oil, e.g. limonene and octanal, in relation to their metabolic pathways[121,122]. Breas et al. highlighted the value of GC-C-IRMS for the rapid and convenient analysis of vanillin and mint hexenol[123]. Karl et al. performed chirospecific measurements of isotope ratios of 2-methylbutanoates from apple flavours[124].

Site specific natural isotope fractionation measured by nuclear magnetic resonance (SNIF-NMR) allows to determine isotope ratios at atomic sites of a substance[38-40,125-126]. This ability can be particularly useful for the authentication of synthetic and natural substances having close or similar molecular isotope values. Indeed, though the molecular isotope value of two substances of similar structure may be equal, their intramolecular distribution of isotopes can be different, owing to their different biosynthetic sources. Danho et al. the reported intramolecular D/H distribution of caffeine from both synthetic and natural origins[127]. Hanneguelle et al. reported strikingly different intramolecular D/H values of linalool from plants and fossil materials[128]. Martin et al. showed that D/H of wine ethanol can be used to sort out varieties according to production and vintage[129]. Martin et al. further carried a systematic investigation of the water cycle in the wine ecosystem[130].

**FORENSIC SCIENCE**

Isotopic analysis can bring information on the geographical origin of drugs[30,31,131-133]. For instance, bearing in mind that heroin is prepared by acetylation of natural morphine, Besacier et al. proposed GC-C-IRMS analysis to locate the geographical origin of heroin and of the acetic anhydride used by the drug trafficker[31].

**HUMIC SUBSTANCES**

Humic substances are complex, partly macromolecular, yellow-brownish organic materials produced during biomass decomposition in soils, in waters and in sediments. Their precise sources, their molecular structure and their transformation pathways are still poorly known[134]. Therefore $^{13}$C analysis and molecular characterisation is a promising approach to study this organic puzzle. By isotopic comparison of various organic components, e.g. linear alkanes, from crop soil and plant Lichtfouse et al. assessed the endogenous origin of humic substances from crop soils[135]. A combined approach using microscopy, pyrolysis, structural characterisation and $^{13}$C/$^{12}$C analysis showed than soil humic substances are formed in part by selective preservation of resistant, aliphatic biopolymers[136-137]. Alternative pathways of humic substances formation were thus proposed[28]. It was also shown on isotopic grounds that linear alkanes from the same soil sample can occur into two compartments, a free, extractable compartment and an older, humin-bound compartment[27,28]. Filley et al. developped $^{13}$C-tetramethylamonium hydroxide (TMAH) chemolysis to track microbial modification of lignin in terrestrial environments[138]. More specifically, $^{13}$C-TMAH thermochemolysis of fungi-degraded lignin followed by gas chromatograhy-mass spectrometry (GC-MS) analysis enabled to determine the amount of hydroxylated aromatic components in degraded lignin residues[138].

**MICROBIOLOGY**

A large number of microbial species living in natural media remain unidentified, notably because they can not be easily grown as pure strains in the laboratory. Here, several recent reports show that investigations coupling $^{13}$C molecular tools and molecular genetic approaches open unexplored research areas[139-141]. Boschker et al. used $^{13}$C labelling of polar lipid fatty acids (PLFAs) to identify the bacteria involved in two important biogeochemical processes: sulphate reduction coupled to oxidation in estuarine sediments, and methane oxidation in a freshwater sediment[139]. Hinrichs et al. further demonstrated the consumption of sedimentary methane by archaebacteria using two lines of evidence: extreme $^{13}$C-depletion of archaebacterial lipid biomarkers (-100‰) and 16S rRNA sequencing[140]. A very promising stable-isotope tool for microbial ecology has also been set up by Radajewski et al.[141]. Indeed, since $^{13}$C-DNA can be separated from $^{12}$C-DNA by centrifugation, it is possible to identify microorganisms living in complex media, e.g. soils, and growing on specific $^{13}$C-labelled sources, e.g. methanol[141]. A such technique should be applied, for instance, to the identification of strains involved in organic pollutant degradation.

**ORGANIC GEOCHEMISTRY**

Following the traditional use of isotopes in geosciences[2], the applications of CG-C-IRMS in biogeochemistry[10] have been blooming early and rapidly, as shown by the special issue edited by Schoell and Hayes[9]. Nonetheless, geochemists have been able to measure isotope values of individual substances[42-57] prior the advent of GC-C-IRMS. Indeed, the fractionation of sedimentary mixtures into pure substances has been performed, for instance by HPLC of amino acids[142,143] and of chlorophyll-derived porphyrins[144-146]. Compound-specific isotope analysis has been applied mainly to source identification, palaeoreconstruction and petroleum geochemistry.

Freeman et al. analysed extracts from the Eocene Messel shale by GC-C-IRMS and found strikingly $^{13}$C-depleted values for bacterial hopanes, suggesting their derivation from methanotrophs living at the time of deposition[147]. Similarly depleted values were observed in the Green River shale by Collister et al.[148], thus allowing to bring some light of the biological sources and food webs of the palaeoenvironment[83]. Lichtfouse and Collister further observed an isotopic correlation between fossil *n*-alkanes and fossil *n*-fatty acids, thus confirming the production of petroleum alkanes by decarboxylation of fatty acids[149]. Collister et al. reviewed *n*-alkane literature and proposed on isotopic grounds various bacterial, planktonic and plant sources for Green River sedimentary *n*-alkanes[150].

Kennicutt and Brooks observed that even-carbon number sedimentary *n*-alkanes are $^{13}$C-enriched relative to their odd-carbon counterparts[151]. By comparison of $^{13}$C/$^{12}$C values of *n*-alkanes from leaf and lake sediments, Rieley inferred the plant origin of long-chain sedimentary *n*-alkanes[152]. Nonetheless other studies suggested possible alternative sources[153,154]. Jones et al. reported $^{13}$C/$^{12}$C values of *n*-alcohols, sterols and an hopanoid alcohol from surface sediments[155]. Kenig et al. presented molecular and isotopic evidence for the presence in sediments of branched alkanes derived from insect waxes[156]. Hartgers et al. assessed the derivation of $^{13}$C-enriched sedimentary carotenoid derivatives from photosynthetic green sulphur bacteria[157] because the biomass of these organisms is isotopically labelled, being enriched in $^{13}$C. Moers et al. observed unusually heavy $^{13}$C/$^{12}$C values (-5‰) of monosaccharides extracted from surface sediments[158].

Organic sulphur is playing a major role in the formation and transformation of sedimentary organic matter[159,160]. Here, isotopic analysis of molecular fossils released by desulfurization from sedimentary macromolecules brought some light on their biological sources, e.g. bacteria, algae, and higher plants[161-165]. In particular, it is shown that free and S-bound substances have different $^{13}$C/$^{12}$C values, suggesting their derivation from different sources.

One of the most fascinating field of application of carbon isotope analysis is the reconstruction of palaeoenvironments, as shown by Schidlowski who inferred life occurrence 3,800 millions years ago using the biosynthetic fingerprint of carbonates versus organic matter[166]. Using $^{13}$C/$^{12}$C ratios of fossil *n*-alkanes, Logan et al. brought some light of the water-column processes ruling the transition from anoxic, bacteria-enriched to oxic, algae-enriched waters during the Proterozoic aeon (2,500-540 Myrs)[167]. The isotopic analysis of fossil photosynthetic pigments revealed the activity of photosynthetic sulphur bacteria in the Black Sea for substantial periods of time in the past, as shown by Sinninghe Damsté et al.[168]. Jasper and Hayes found that $^{13}$C/$^{12}$C ratios of fossil algal alkenones recorded atmospheric $CO_2$ levels during the late quaternary (0-100 kyrs)[169]. Freeman and Hayes explained the principles of calculation of ancient $CO_2$ levels, monitoring for example the $^{13}C/^{12}C$ values of porphyrins, the chlorophyll fossils[170]. In order to further understand ancient biogeochemical cycles, investigations of modern water-column processes are under way[171].

Petroleum geochemists have extensively used carbon isotopes[172-175] because the processes implied in the formation and migration of petroleum in sedimentary rocks are still obscure. Bjorøy et al. found clear $^{13}C/^{12}C$ differences for *n*-alkanes versus acyclic isoprenoids in North Sea oils[176]. Schoell et al. showed that fossil biomarkers from Miocene tar-like petroleum can be related to their biological sources according to their $^{13}C/^{12}C$[177]. Lichtfouse and Collister inferred the formation of petroleum *n*-alkanes by decarboxylation of *n*-fatty acids[149]. Gõni and Eglinton investigated the $^{13}C/^{12}C$ ratios of aliphatic and phenolic substances released by pyrolysis of petroleum precursors[178]. Simoneit and Schoell confirmed the generation of PAHs from hydrothermal petroleum by breakdown of sedimentary kerogen, the macromolecular fraction of sedimentary organic matter[179]. Lichtfouse et al. found that $^{13}C/^{12}C$ ratios of *n*-alkanes of deep rocks and petroleums from Paris basin fall in a narrow range[180], suggesting that the carbon of only a restricted number of living organisms is contributing to petroleum genesis.

By setting up $^{14}C$ dating of trace organics at the molecular level using preparative GC, Eglinton et al. have opened a novel vista in organic geochemistry[60,61]. Indeed, the assessment of relationships between molecules occurring in complex media such as soils and sediments have so far mainly relied on structural grounds. Now, $^{14}C$ dating provides an independent way to unravel the biological sources of sedimentary organic molecules. For instance, in a 4-7 cm marine sediment of bulk organic matter age of 880 years BP, the large age difference observed between the $C_{25}$ *n*-alkane (1340 years) and the $C_{31}$ *n*-alkane (500 years) can be explained by a higher contribution of fossil carbon in the former[61]. These results are in agreement with the fossil fuel origin of short-chain linear alkanes found in soils[89]. It is thus expected that $^{14}C$ molecular dating will find applications into various fields, notably for environmental issues because human activities have introduced large amounts of fossil fuel $^{14}C$-(almost)free carbon in ecosystems (pollutants, plastics, solvents, drugs, etc.).

**SOIL SCIENCE**

Soil organic matter has many agronomic functions such as water retention, plant-nutrient sequestration, and holding of minerals[91,181]. As in any complex ecosystem, the behaviour of soil organic molecules is far from being understood. Most soil organic substances still remain unidentified. Here, isotopic studies at the molecular level appear as a promising way to unravel the multiple pathways of organic matter transformation. For instance, the long-term turnover of plant wax *n*-alkanes[17] in soil has been monitored using a 23-year natural labelling experiment by cultivation of maize[19,20]. The plant origin of long-chain *n*-alkanes[182], *n*-alcohols[82] and *n*-fatty acids[82], and the microbial origin of short-chain fatty acids[75] have been assessed using both natural and artificial $^{13}C$-labelling experiments. Huang et al. studied the $^{13}C/^{12}C$ values of lipids of stratified organic soils[183]. They found that hopanoid were $^{13}$C-enriched by 4‰ relative to plant *n*-alkanes, suggesting a source of heterotrophic bacteria feeding on carbohydrates or proteins. A 23-year plant decomposition experiment further revealed the absence of isotopic fractionation of *n*-alkanes, thus validating their use as proxies for palaeoenvironmental reconstruction and as tracers of carbon flow in soils and sediments[184].

**SPORT**

The detection of drugs used to improve sport performances is difficult because administrated substances are usually occurring at trace levels in urine and in blood. It is even more difficult when the drugs, e.g. steroids, are already naturally biosynthesized by the body. Here, similarly to aroma authentication, natural and synthetic homologues can be isotopically distinguished because the carbon of the synthetic substance is of petroleum origin. For instance, the $^{13}$C/$^{12}$C analysis of urinary steroids to assess the misuse of testosterone has been set up[15,185-189]. Aguilera et al. further extended this technique to the detection of hydrocortisone used for horseracing[190].

**CONCLUSION**

The recent development of mass spectrometric and NMR techniques allowing the determination of isotopic composition at the molecular and atomic levels has opened novel research fields into various scientific disciplines. Such techniques are already used for industrial applications, e.g. petroleum research and food autentication. Since there is a growing need for tracers to resolve various issues related to complex systems, it is expected that isotopic analysis at the molecular and atomic levels will become widely used. $^{14}$C-molecular dating, GC-C-IRMS ($^{13}$C, $^2$H, $^{15}$N), and intramolecular isotopic analysis by SNIF-NMR look very promising. Molecular investigations undertaken at the frontiers of several fields, e.g. chemistry, geology and biology[191], could be particurlarly fruitful.

**ACKNOWLEDGEMENTS**

This article is dedicated to John Hayes, Dawn Merritt, Kate Freeman, Jim Collister, David Hollander, Glenn Hieshima, John Jasper, Lisa Pratt, John Rupp and all other fellows from Indiana University, Bloommington IN, with whom I have had a good time and hard work during my post-doctoral stay in 1990. I also thank Dr. Nick Ostle, Institute of Terrestrial Ecology for his kind invitation to present a conference on CSIA at the annual meeting of the stable isotope mass spectrometry users group (SIMSUG), Grange-Over-Sands, UK (18th-19th January 2000).